\documentclass[preprints,article,accept,moreauthors,pdftex]{Definitions/mdpi}

\firstpage{1}
\pubvolume{1}
\issuenum{1}
\articlenumber{0}
\pubyear{2022}
\copyrightyear{2022}
\externaleditor{{Academic Editors: Liwei Geng and Yongke Yan} }
\datereceived{19 May 2022}
\dateaccepted{27 June 2022 }
\datepublished{}
\hreflink{https://doi.org/} % If needed use \linebreak
\Title{Magnetic Collapse in Fe$_3$Se$_4$ under High Pressure}

\TitleCitation{Magnetic Collapse in Fe$_3$Se$_4$ under High Pressure}

\Author{Lyudmila  V. Begunovich $^{1,2}$\orcidA{}, Maxim M. Korshunov $^{1,2,}$*\orcidB{}  and Sergey G. Ovchinnikov $^{1,2}$\orcidC{}}

\AuthorNames{ Lyudmila V. Begunovich,  Maxim M. Korshunov  and  Sergey  G. Ovchinnikov}

\AuthorCitation{Begunovich, L.V.; Korshunov, M.M.; Ovchinnikov, S.G.}
\address{%
$^{1}$ \quad Kirensky Institute of Physics, Federal Research Center KSC SB RAS, Akademgorodok 50/38,  \mbox{660036 Krasnoyarsk, Russia;} lyuda.illuzia@gmail.com (L.V.B.); sgo@iph.krasn.ru (S.G.O.) \\
$^{2}$ \quad Siberian Federal University, Svobodny Prospect 79,  660041 Krasnoyarsk, Russia}

\corres{Correspondence: mkor@iph.krasn.ru}

\abstract{Electronic structure and magnetic properties of Fe$_3$Se$_4$ are calculated using the density functional approach. Due to the metallic properties, magnetic moments of the iron atoms in two nonequivalent positions in the unit cell are different from ionic values for Fe$^{3+}$ and Fe$^{2+}$ and are equal to $M_1=2.071 \mu_B$ and $M_2=-2.042 \mu_B$, making the system ferrimagnetic. The total magnetic moment for the unit cell is $2.135 \mu_B$. Under isotropic compression, the total magnetic moment decreases non-monotonically and correlates with the non-monotonic dependence of the density of states at the Fermi level $N(E_F)$. For 7\% compression, the magnetic order changes from the ferrimagnetic to the ferromagnetic. At 14\% compression, the magnetic order disappears and the {total} magnetic moment becomes zero, {leaving the system in a paramagnetic state}. This compression corresponds to the pressure of 114~GPa. {The magnetic ordering changes faster upon application of an isotropic external pressure due to the sizeable anisotropy of the chemical bondings in Fe$_3$Se$_4$. The ferrimagnetic and paramagnetic states occur under pressures of 5.0~and 8.0~GPa, respectively.} The system remains in the metallic state for all values of compression.}


\begin{document}
%%%%%%%%%%%%%%%%%%%%%%%%%%%%%%%%%%%%%%%%%%

%%%%%%%%%%%%%%%%%%%%%%%%%%%%%%%%%%%%%%%%%%
\section{Introduction}

Magnetic collapse associated with the disappearance of magnetic moments in $3d$ ions is observed in many insulating transition metal oxides (Mn--O, Fe--O, Co--O and Ni--O systems). A review of experimental data for iron oxides~\cite{Lyubutin2009} revealed that the spin crossover between the high-spin and low-spin states of the cation in most cases takes place with increasing pressure and critical pressure is close to 50--70~GPa. The crystals with Fe$^{2+}$ ions have a low-spin state with a zero spin ($S=0$) that leads to the appearance of a nonmagnetic phase. In the case of crystals with Fe$^{3+}$ ions, the low-spin state exhibit $S=1/2$, so the magnetic state can be preserved, albeit at a lower critical temperature, such as in FeBO$_3$~\cite{Gavriliuk2005}. For magnetite Fe$_3$O$_4$ containing both Fe$^{2+}$ and Fe$^{3+}$ ions, a new nonmagnetic phase was experimentally found~\cite{Mao1974,Pasternak1994} at pressures above 25~GPa and room temperature.

The iron selenides FeSe$_x$ ($1 \leq x \leq 1.33$) form phases with iron vacancies that crystallize into structures derived from the hexagonal NiAs-type structure. Among them, two compounds, Fe$_7$Se$_8$ and Fe$_3$Se$_4$, have superstructures with ordered Fe vacancies~\cite{Okazaki1956}. Fe$_7$Se$_8$ has a hexagonal structure whereas Fe$_3$Se$_4$ has a monoclinic structure isomorphic to Cr$_3$X$_4$ (X = S, Se, Te). In an Fe$_3$Se$_4$ unit cell, the vacancies of Fe appear in every second iron layer. The presence of ordered Fe vacancies facilitates the appearance of the ferrimagnetic state in Fe$_3$Se$_4$~\cite{Tewari2020,Lambert-Andron1969,Adachi1961,Pohjonen2018}. The experimental values of the total magnetic moment are in the range from $0.69 \mu_B$ to $1.17 \mu_B$ per formula unit (f.u.)~\cite{Tewari2020,Lambert-Andron1969,Adachi1961}. The magnetic moments on Fe ions estimated within the framework of the ionic model~\cite{Lambert-Andron1969} are too large ($3.25 \mu_B$ for site 1 and $1.94 \mu_B$ for site 2) and are not validated by experimental data~\cite{Lambert-Andron1969,Pohjonen2018,Andresen1968}. Neutron diffraction on Fe$_3$Se$_4$ gives smaller effective spin values for two Fe positions, $S_1 = 1.08 \mu_B$ and $S_2 = 0.71 \mu_B$~\cite{Andresen1968}. In addition, a study by M\"{o}ssbauer revealed the very low average internal magnetic fields for Fe sites~\cite{Pohjonen2018}, which do not correspond to regular high-spin Fe$^{3+}$ or Fe$^{2+}$ states. All of the abovementioned studies indicate the delocalization of $3d$ electrons of iron ions and the inapplicability of the model of localized magnetic moments to the system.

In the series Fe$_3$O$_4$--Fe$_3$S$_4$--Fe$_3$Se$_4$, on the one hand, the presence of two non-equivalent positions of cations and ferrimagnetic properties are preserved, on the other hand, an increase in covalence enhances the metallic properties. Despite the structural similarity between greigite Fe$_3$S$_4$ and magnetite Fe$_3$O$_4$, their magnetic properties~\cite{Chang2008} and magneto-optical spectra~\cite{Lyubutin2013} differ significantly. The differences from magnetite become even greater for Fe$_3$Se$_4$, which has interesting magnetic and electrical properties~\cite{Adachi1961,Pohjonen2018,Andresen1968}. Fe$_3$Se$_4$ is a metallic ferrimagnetic material. Its electronic structure was calculated within the density functional theory (DFT)~\cite{Singh2020}. Tewari et al.~\cite{Tewari2020} declare that the Fe$_3$Se$_4$ material possess half-metallic properties. The spin-down band gap ($E_g$) and half-metallic energy gap ($E_{HM}$) calculated using the HSE06 hybrid functional~\cite{Krukau2006} were found to be 1.8~eV and 0.17~eV, respectively. However, the analysis of experimental data reveled the extremely small $E_{HM}$ (1.3~meV and 34~meV, depending on the Se stoichiometry), thus the low energy is required to occupy minority spin states by the majority spin carriers during spin flip processes. The discrepancy between the theoretical and experimental $E_{HM}$ values can be associated with an increase in the localization of electronic states when a fraction of the Hartree--Fock exchange energy is included. This results in the inaccurate description of electronic structure near the Fermi level. It is known that the HSE06 functional can overestimate band gap~\cite{Meng2016,Noh2014}. Moreover, Gao et al.~\cite{Gao2016} suggest that hybrid functionals, such as HSE and PBE0 are not suitable for studying metal systems, and the LDA or GGA give better results in describing metallic properties.

Here we study the properties of Fe$_3$Se$_4$ under high pressure within DFT. We calculated the changes of the magnetic moment on each sublattice with the increasing isotropic compression. The values of the magnetic moments decrease non-monotonically and eventually vanish for both inequivalent positions. Since Fe$_3$Se$_4$ possess metallic properties, such a magnetic collapse cannot be represented as the energy-level crossing of the high-spin and the low-spin cation states. In this case, the magnetic collapse is caused by the alignment of the numbers of spin-up and spin-down electrons on each cation such that one can call this the itinerant analogue of the spin crossover. {The spin crossover occurs faster under compression by isotropic pressure compared to when compression by isotropic strain.}

%%%%%%%%%%%%%%%%%%%%%%%%%%%%%%%%%%%%%%%%%%
\section{Computational Details}

Calculations of atomic and electronic structure and magnetic properties were performed in the framework of density functional theory using the Vienna ab initio simulation package (VASP) \cite{Kresse1996}. Exchange-correlation effects were described by the Perdew--Burke--Ernzerh (PBE) of generalized gradient approximation (GGA)~\cite{Perdew1996}. The ion-electron interactions were represented by the projector-augmented wave method (PAW)~\cite{Blochl1994}, and the plane-wave cutoff energy of 600~eV was applied. The criteria for the total energy minimization and interatomic forces were set to $10^{-4}$~eV and $10^{-2}$~eV/\AA, respectively. The energy convergence criteria were decreased to $10^{-5}$~eV to obtain a more accurate electronic structure. The first Brillouin zone (1BZ) was sampled by $24 \times 14 \times 8$ grid using the Monkhorst--Pack scheme~\cite{Monkhorst1976}. {The isotropic compressive strain was modelled as the change of the structural parameters. The value of strain is defined as $\xi = (l_0-l)/l_0$, where $l_0$ and $l$ are the equilibrium and strained lattice constant, respectively. The strain was uniform on three lattice directions, i.e., all lattice constants $a$, $b$, and $c$ were changed at the same time by the same value of $\xi$. The strained lattice constants did not change during the geometry optimization. Compressive strain ranging from 0 to 14\% is considered. The lattice shape remains unchanged during the compression. Compression by an isotropic pressure was studied by adding the external pressure to the diagonals of stress tensors. The value of the external pressure was changed from zero up to 8~GPa. Lattice constants, cell shape, cell volume, and ionic positions were optimized for these structures.} The Visualization for Electronic and Structural Analysis (VESTA)~\cite{Momma2011} software was used for representation of atomic structures. The ``vaspkit'' software~\cite{Wang2021} was used for post-processing.

%%%%%%%%%%%%%%%%%%%%%%%%%%%%%%%%%%%%%%%%%%
\section{Results and Discussion}

\begin{figure}[H]

\includegraphics[width=0.45\linewidth]{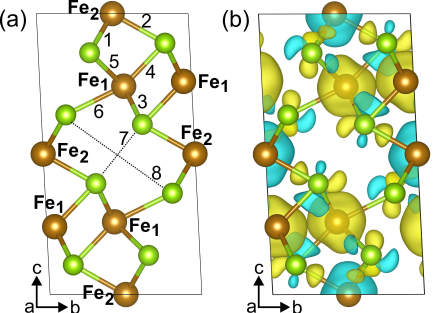}
\caption{Atomic structure (\textbf{a}) and {the spin-resolved magnetization density (\textbf{b}) of Fe$_3$Se$_4$. The unit cell is marked with black lines. Brown and yellow-green colors correspond to Fe and Se atoms, respectively. In panel (\textbf{b}), yellow (blue) areas indicate the spin-up (spin-down) density. Isosurface value is 0.002 $a_0^{-3}$,where $a_0$ is the Bohr radius.}  \label{fig:StructSpin}}
\end{figure}

\begin{figure}[H]

\includegraphics[width=0.98\linewidth]{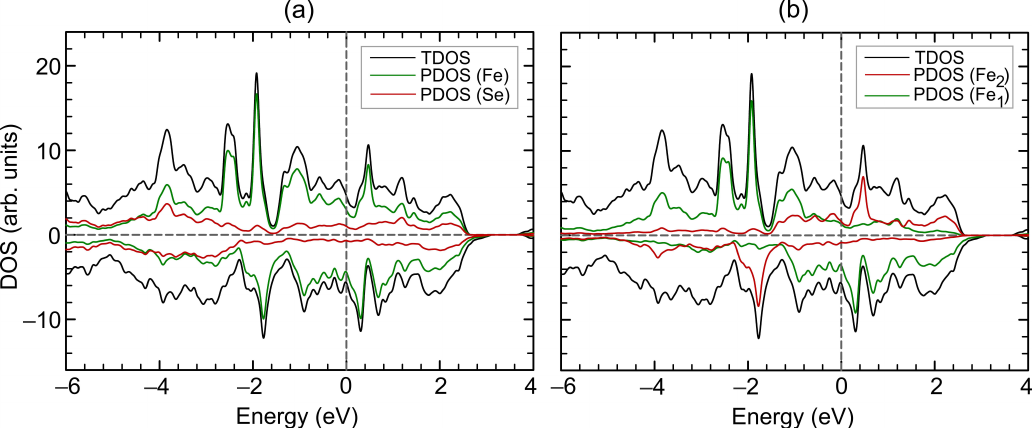}
\caption{Density of states (\textbf{a},\textbf{b}) of Fe$_3$Se$_4$. Total DOS of Fe$_3$Se$_4$ is shown by black curves. In panel (\textbf{a}), partial DOS of Fe and Se atoms are shown by green and red curves, respectively. In panel (\textbf{b}), partial DOS of iron atoms on site 1 and on site 2 are shown by green and red curves, respectively. Positive and negative values of DOS corresponds to spin-up and spin-down channels, respectively. The Fermi level corresponds to zero. \label{fig:DOS}}
\end{figure}

\begin{figure}[H]

\includegraphics[width=0.58\linewidth]{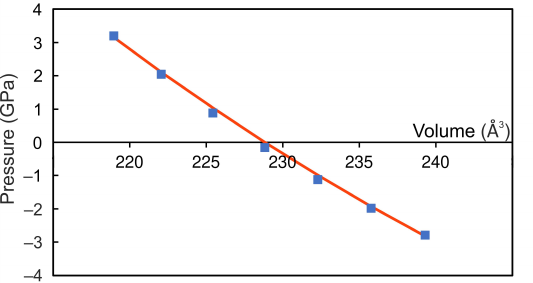}
\caption{Pressure as a function of Fe$_3$Se$_4$ volume. Blue dots indicate values for structures with finite strains $\pm 0.5$, $\pm 1.0$, $\pm 1.5$\%. The curve through this values fitted to the Birch–Murnaghan equation is marked by red. \label{fig:PV}}
\end{figure}

The crystal structure of the bulk Fe$_3$Se$_4$ is shown in Figure~\ref{fig:StructSpin}a. The unit cell relates to the $I2/m$ space group with the structural parameters $a=6.071$~\AA, $b=3.377$~\AA, $c=11.174$~\AA, and $\beta=92.818^\circ$, which are well in agreement with previously reported experimental data~\cite{Tewari2020,Pohjonen2018,Andresen1968}, see Table~\ref{tab:struct} for comparison. The unit cell contains six Fe atoms and eight Se atoms, with iron atoms occupying two non-equivalent positions, namely, octahedral (Fe$_2$) and distorted-octahedral (Fe$_1$) sites. {The Se$-$Fe$_2-$Se angles for distorted-octahedral sites are equal to 104.0$^\circ$, 89.8$^\circ$, 89.9$^\circ$ and 75.5$^\circ$.} The bond lengths between Fe$_2$ and surrounding Se ions are 2.468~\AA~and 2.413~\AA~{(bonds 1 and 2 in Figure~\ref{fig:StructSpin}a, respectively)} and Fe$_1$--Se bond lengths are 2.390~\AA, 2.374~\AA, 2.545~\AA, and 2.581~\AA~{(bond 3, 4, 5, and 6 in Figure~\ref{fig:StructSpin}a, respectively)}, see Table~\ref{tab:bonds}. The distances between Se atoms closest to the iron vacancy are 2.996~\AA~and 5.277~\AA. The magnetic structure was found to be ferrimagnetic. {The coupling between spins in Fe$_3$Se$_4$ is accomplished by the mechanism of ``through-bond spin polarization'' that is a specific realization of the superexchange cation--anion--cation interaction, i.e., an iron atom with spin-up (spin-down) density induces spin-down (spin-up) density on the $p$-orbital of the adjacent Se atom directly bonded to it. Thus an iron atom is bonded to a selenium atom through the $p$-orbital with the direction of the electron spin opposite to its spin density. The distribution of the magnetisation density is shown in Figure~\ref{fig:StructSpin}b.} The calculated magnetic moments on iron in sites 1 and 2 are $2.071 \mu_B$ and $-2.042 \mu_B$, respectively. {The magnetic moments on Se atoms are very small and equal to $-0.018 \mu_B$ and $0.015 \mu_B$.} The total magnetic moment of Fe$_3$Se$_4$ is $2.135 \mu_B$/f.u., i.e., in the range of the previously DFT-calculated values~\cite{Tewari2020,Singh2020,Long2011} but larger than the known experimental values~\cite{Tewari2020,Lambert-Andron1969,Adachi1961}, see Table~\ref{tab:struct}. The discrepancy in magnetic moment between experimental and DFT-predicted values relates to the delocalization of $3d$ electrons of Fe ions. Our calculations showed that the electronic structure of Fe$_3$Se$_4$ is metallic, see Figure~\ref{fig:DOS}a,b, as in ~\cite{Singh2020,Persson2016}, and half-metallic state was not found. The density of states (DOS) at the Fermi level are predominantly formed by iron states. The bulk modulus of Fe$_3$Se$_4$ is determined by performing six finite distortions of the lattice with $\pm 0.5$, $\pm 1.0$, $\pm 1.5$\% magnitude. The calculated $P-V$ data are fitted to the Birch–Murnaghan equation of state $P = 3/2 B_0 \left[u^{-7⁄3}-u^{-5⁄3}\right] \cdot \left[1-3/4 (4-B')(u^{-2⁄3}-1)\right]$, where $B_0$, $u=V⁄V_0$, $V_0$, $V$, and $B'$ are the bulk module, the dimensionless volume, the reference volume (the initial volume of Fe$_3$Se$_4$), the deformed volume, and the derivative of the bulk modulus with respect to pressure, respectively. Bulk modulus as a function of volume is shown at Figure~\ref{fig:PV}. The obtained value of $B_0$ is 68 GPa, which is very similar to that of Fe$_3$S$_4$ (\mbox{62.8 GPa})~\cite{Roldan2013} and is in the range of values for isomorphic Cr$_3$Se$_4$ (57.7 GPa) and Cr$_3$S$_4$ (72.9 GPa)~\cite{Guo2009} structures. Its pressure derivative was found to be six.

\begin{table}[H]
\caption{Structural and magnetic parameters of monoclinic phase of Fe$_3$Se$_4$. \label{tab:struct}}
\newcolumntype{C}{>{\centering\arraybackslash}X}
\begin{tabularx}{\textwidth}{CCCCCC}
\toprule
\multicolumn{4}{c}{\textbf{Structural Parameters}} & \textbf{Magnetic} &  \textbf{Ref.}\\
\cline{1-4}
\multicolumn{3}{c}{\textbf{Lattice Constant (\AA)}} & \textbf{Angle (\boldmath{$^\circ$})} & \textbf{Moment} & \\
\cline{1-4}
\boldmath{$a$} & \boldmath{$b$} & \boldmath{$c$} & \boldmath{$\beta$} & \textbf{(\boldmath{$\mu_B$}/f.u.)}& \\
\midrule
6.202 & 3.532 & 11.331 & 91.825$^\circ$ & - & \cite{Pohjonen2018}\\
6.113 & 3.486 & 11.139 & 91.66$^\circ$ & - & \cite{Andresen1968}\\
6.16 & 3.53 & 11.10 & 92.0$^\circ$ & - & \cite{Okazaki1956}\\
- & - & - & - & 1.17 & \cite{Tewari2020}\\
- & - & - & - & 0.9 & \cite{Lambert-Andron1969}\\
- & - & - & - & 0.69 & \cite{Adachi1961}\\
6.071 & 3.377 & 11.174 & 92.818$^\circ$ & 2.128 & This work\\
\bottomrule
\end{tabularx}
\end{table}\vspace{-6pt}

\begin{table}[H]
\caption{{Bond lengths (\AA) and distance (\AA) between Se atoms closest to the iron vacancy of the Fe$_3$Se$_4$ monoclinic phase. The atomic numbering scheme is shown in Figure~\ref{fig:StructSpin}a.} \label{tab:bonds}}
\newcolumntype{C}{>{\centering\arraybackslash}X}
\begin{tabularx}{\textwidth}{m{2cm}<{\centering}CCCCCCCC}
\toprule
 & \multicolumn{8}{c}{\textbf{Bond or distance (\AA)}}\\
\cline{2-9}
\textbf{Strain (\%)} & \textbf{1} & \textbf{2} & \textbf{3} & \textbf{4} & \textbf{5} & \textbf{6} & \textbf{7} & \textbf{8}\\
\midrule
0 & 2.468 & 2.413 & 2.390 & 2.374 & 2.545 & 2.581 & 2.996 & 5.277\\
2 & 2.419 & 2.364 & 2.349 & 2.355 & 2.503 & 2.516 & 2.895 & 5.151\\
3 & 2.396 & 2.342 & 2.325 & 2.344 & 2.480 & 2.481 & 2.849 & 5.091\\
3.5 & 2.384 & 2.333 & 2.312 & 2.338 & 2.467 & 2.465 & 2.828 & 5.066\\
4 & 2.381 & 2.335 & 2.291 & 2.324 & 2.440 & 2.446 & 2.825 & 5.068\\
4.5 & 2.370 & 2.319 & 2.283 & 2.323 & 2.426 & 2.428 & 2.796 & 5.044\\
5 & 2.346 & 2.301 & 2.281 & 2.316 & 2.424 & 2.429 & 2.756 & 4.994\\
5.5 & 2.338 & 2.280 & 2.268 & 2.330 & 2.416 & 2.393 & 2.729 & 4.967\\
6 & 2.325 & 2.276 & 2.264 & 2.305 & 2.397 & 2.400 & 2.700 & 4.943\\
6.5 & 2.314 & 2.262 & 2.253 & 2.303 & 2.387 & 2.382 & 2.671 & 4.916\\
7 & 2.306 & 2.245 & 2.239 & 2.311 & 2.379 & 2.352 & 2.645 & 4.888\\
8 & 2.282 & 2.220 & 2.220 & 2.298 & 2.358 & 2.327 & 2.588 & 4.829\\
9 & 2.258 & 2.196 & 2.202 & 2.283 & 2.337 & 2.301 & 2.535 & 4.767\\
10 & 2.235 & 2.172 & 2.183 & 2.266 & 2.315 & 2.274 & 2.483 & 4.705\\
11 & 2.212 & 2.147 & 2.164 & 2.250 & 2.294 & 2.248 & 2.432 & 4.644\\
12 & 2.189 & 2.123 & 2.145 & 2.231 & 2.272 & 2.222 & 2.385 & 4.582\\
13 & 2.166 & 2.100 & 2.127 & 2.211 & 2.250 & 2.195 & 2.337 & 4.519\\
14 & 2.143 & 2.078 & 2.109 & 2.189 & 2.277 & 2.169 & 2.292 & 4.455\\
\bottomrule
\end{tabularx}
\end{table}\vspace{-9pt}

\vspace{-6pt}

\begin{figure}[H]

\includegraphics[width=0.6\linewidth]{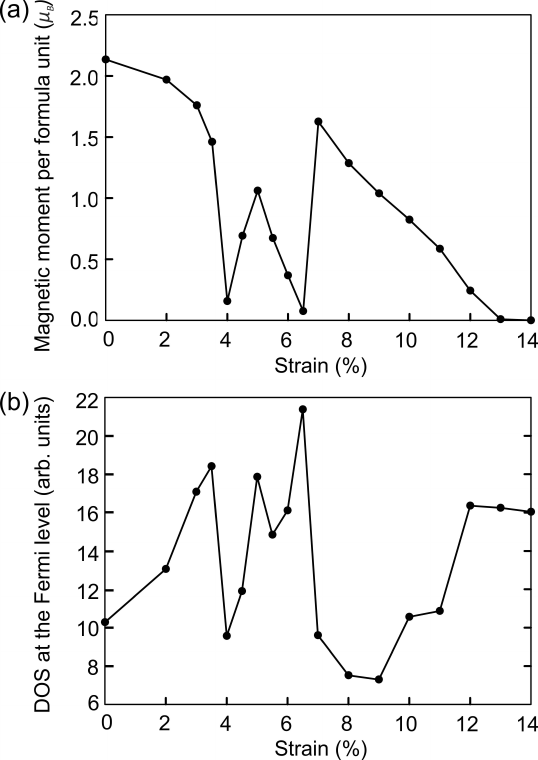}
\caption{The dependence  of the total magnetic moment (\textbf{a}) and DOS at the Fermi level (\textbf{b}) on isotropic compression. \label{fig:DOSMagMom}}
\end{figure}

Next we study the effect of strain on the magnetic and electronic properties of Fe$_3$Se$_4$. First approach is to model the isotropic compressive strain via the change of the structural parameters.
{The lattice of Fe$_3$Se$_4$ remains monoclinic and the angles between vectors remain unchanged during compression.} Isotropic compression along the lattice constant up to 4\% reduces the total magnetic moment of the system to $0.157 \mu_B$/f.u., see Figure~\ref{fig:DOSMagMom}a and Table~\ref{tab:moments}, {which is caused by a decrease in bond lengths, see Table~\ref{tab:bonds}. The reduction in bond lengths makes the covalent character greater than the ionic character, which causes a decrease in the spin polarization of atoms.} Further compression to 5\% increases the magnetic moment by $0.904 \mu_B$/f.u. {In this compression range, the bond lengths with iron atoms located in the vacant layer (Fe$_2-$Se) decrease faster (by near 1.5\%) than Fe$_1-$Se bond lengths (by 0.44, 0.34, 0.66, 0.70\%). This leads to the fact that the magnetic moment on the Fe$_2$ atoms drastically decreases by 65.4\% upon compression from 4 to 5\%. At the same time, the magnetic moment on Fe$_1$ atoms increases from 0.602 to 0.764 $\mu_B$. A significant decrease in the magnetization on Fe$_2$, opposite in direction to the magnetization of Fe$_1$, leads to an increase in the total magnetic moment in the system. With further compression, the change in the Fe$_2-$Se bond lengths slows down and the change in Fe$_1-$Se bond lengths increases, causing} the magnetic moment to decrease to $0.124 \mu_B$/f.u. (6.5\% compression).

\begin{table}[H]
\caption{Magnetic moments (in~$\mu_B$) of iron atoms on two sites, Fe$_1$ and Fe$_2$, and total magnetic moments per Fe$_3$Se$_4$ formula unit of Fe$_3$Se$_4$ under compression. {The Wigner--Seitz radius for Fe ions is 1.302 \AA}  \label{tab:moments} }
\newcolumntype{C}{>{\centering\arraybackslash}X}
\begin{tabularx}{\textwidth}{CCCC}
\toprule
\textbf{Strain (\%)} & \textbf{Fe\boldmath{$_1$} Moment} & \textbf{Fe\boldmath{$_2$} Moment} & \textbf{Total Moment}\\
\midrule
0 & 2.071 & $-$2.042 & 2.135\\
2 & 1.778 & $-$1.631 & 1.969\\
3 & 1.556 & $-$1.417 & 1.759\\
3.5 & 1.387 & $-$1.318 & 1.460\\
4 & 0.602 & $-$1.334 & 0.157\\
4.5 & 0.282 & $-$1.214 & 0.692\\
5 & 0.764 & $-$0.462 & 1.061\\
5.5 & 0.493 & $-$0.297 & 0.673\\
6 & 0.266 & $-$0.209 & 0.368\\
6.5 & 0.053 & $-$0.041 & 0.075\\
7 & 0.463 & 0.666 & 1.626\\
8 & 0.352 & 0.540 & 1.286\\
9 & 0.282 & 0.425 & 1.038\\
10 & 0.225 & 0.328 & 0.823\\
11 & 0.162 & 0.230 & 0.585\\
12 & 0.068 & 0.094 & 0.243\\
13 & 0.004 & 0.005 & 0.009\\
14 & 0.000 & 0.000 & 0.000\\
\bottomrule
\end{tabularx}
\end{table}\vspace{-6pt}

Compression to 7\% leads to a sharp jump in the magnetic moment up to $1.626 \mu_B$ and the magnetic order changes from ferrimagnetic to ferromagnetic. {The magnetic moments on Fe$_1$ and Fe$_2$ ions are $2.071 \mu_B$ and $-2.042 \mu_B$, respectively. The magnetic moments on Se atoms are practically absent and equal to $0.006 \mu_B$.} The bond lengths of Fe$_3$Se$_4$ under 7\% strain is unevenly decreased by 2.26--8.87\%. The distances between selenium atoms located near the iron vacancy are reduced by 7.4 and 11.7\% compared to the initial distance, and equal to 4.888~\AA~and 2.645~\AA, Table~\ref{tab:bonds}. The magnetic moment decreases linearly with the further compression from 7\% to 13\% and vanishes under 14\% of compression. The pressure corresponding to this strain is equal to 114~GPa and the volume of the cell is 145.55~\AA$^3$. In this structure Fe--Se bond lengths are 7.79--15.96\% shorter compared to the bond lengths in the original structure. The distances between selenium atoms near the vacancy decrease by 15.6 and 23.5\% and are equal to 4.455~\AA~and 2.292~\AA. {Octahedrons formed by Fe$_2$-Se bonds are disodered, so that Se$-$Fe$_2-$Se angles are equal to 85$^\circ$~and~95$^\circ$. Selenium atoms located at opposite corners of the octahedron are still in the same plane (Se$-$Fe$_2-$Se angles are 180$^\circ$). The Se$-$Fe$_2-$Se angles are equal to 108.3$^\circ$, 89.5$^\circ$, 86.6$^\circ$, and 74.7$^\circ$.} Values of the iron magnetic moments on sites 1 and 2 are shown in Table~\ref{tab:moments}.

\begin{figure}[H]

\includegraphics[width=1.0\linewidth]{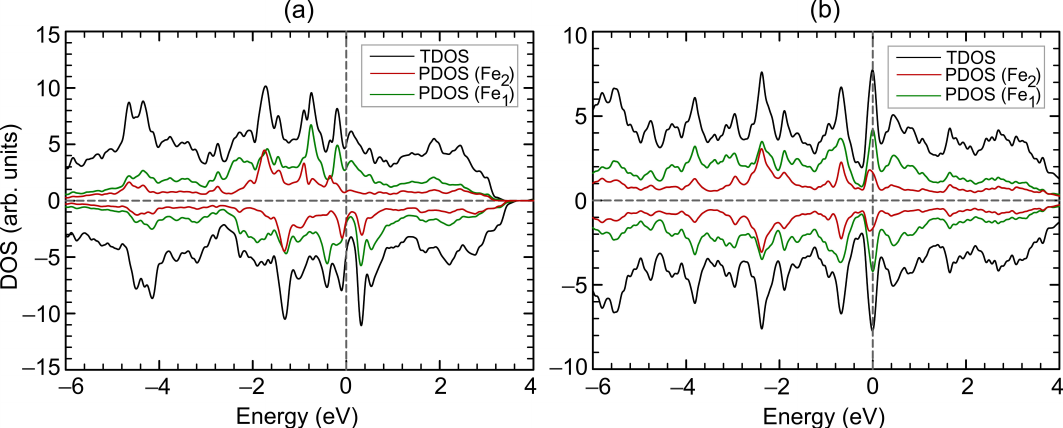}
\caption{Total (TDOS) and partial (PDOS) densities of states for Fe$_3$Se$_4$ under 7\% (\textbf{a}) and \mbox{14\% (\textbf{b})} compression. Positive and negative values corresponds to spin-up and spin-down channels, respectively. The Fermi level corresponds to zero. \label{fig:DOSm7m14}}
\end{figure}

The electronic structure of Fe$_3$Se$_4$ remains metallic throughout the studied compression range. The DOS at the Fermi level $N(E_F)$ as a function of strain is presented at Figure~\ref{fig:DOSMagMom}b. Slight compression up to 3.5\% leads to an increase in the density of states at the Fermi level. Further, the $N(E_F)$ non-monotonically depends on the strain and correlates with the non-monotonic dependance of the total magnetic moment. An increase and decrease in the total magnetic moment is accompanied by an increase and decrease in the $N(E_F)$. The decrease in the magnetic moment during compression from 8 to 14\% is accompanied by an increase in the $N(E_F)$. Figure~\ref{fig:DOSm7m14} shows DOS of Fe$_3$Se$_4$ at critical points. The redistribution of DOSs are observed, compared with the DOS of the original structure (Figure~\ref{fig:DOS}b). Compression up to 7\% results in $N(E_F)$ decreasing, and the vacant states shifting by $0.6$~eV to higher energies. The density of states with spin-up and spin-down is equalized on each Fe cation in the structure at a deformation of 14\% (Figure~\ref{fig:DOSm7m14}b), which leads to the disappearance of the magnetic order. It is the itinerant analogue of the spin crossover in the metallic system. The metallic system of itinerant electrons lacks for the long-range magnetic order thus ending in the Pauli paramagnetic state.
In this case, the DOS at the Fermi level is increased. The occupied  states decrease significantly at $-0.25$~eV and increase at lower energies. The vacant states decrease at $0.19$~eV and increase at higher energies.

\begin{figure}[H]

\includegraphics[width=0.5\linewidth]{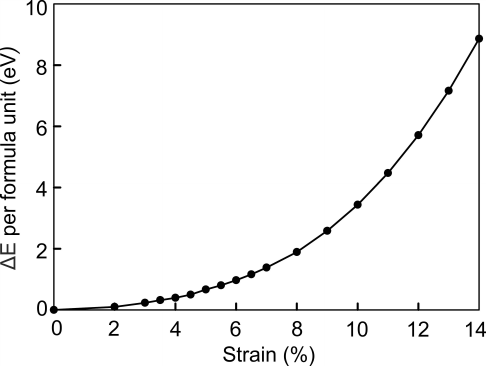}
\caption{{Energetic stability of the compressed Fe$_3$Se$_4$ structure relative to the original structure, $\Delta E$ in eV per Fe$_3$Se$_4$ formula unit.} \label{fig:Energy}}
\end{figure}

{The energy difference $\Delta E$ per formula unit between compressed and original Fe$_3$Se$_4$ was calculated as $\Delta E = E_{compressed} - E_{original}$. Here $E_{compressed}$ and $E_{original}$ are the total energy of the Fe$_3$Se$_4$ cell under isotropic compression and the total energy of the non-compressed Fe$_3$Se$_4$ cell, respectively. The dependence of $\Delta E$ on compressive strain is shown in Figure~\ref{fig:Energy}. The increase in the energy is described by the cubic polynomial $f=0.0032x^3-0.0048x^2+0.0759x-0.0241$. The structure at critical points of 7 and 14\% is higher in energy by $1.38$ and $8.86$~eV/f.u., respectively.}

\begin{figure}[H]

\includegraphics[width=0.74\linewidth]{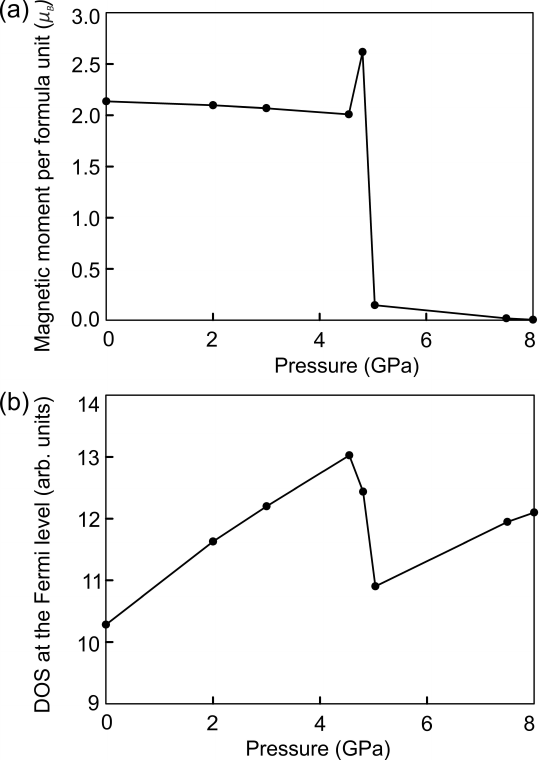}
\caption{{The dependence of the total magnetic moment (\textbf{a}) and DOS at the Fermi level (\textbf{b}) on \mbox{external pressure.}} \label{fig:DOSMagMomPstress}}
\end{figure}

{Second approach to studying the effect of compression of Fe$_3$Se$_4$ cells is to apply the isotropic pressure by adding the external pressure to the diagonals of the stress tensors. The results of calculations show that the cell shape of Fe$_3$Se$_4$ does not change over the entire interval of the external pressure. Lattice constants vary nonuniformly because the cell is anisotropic, see Table~\ref{tab:latticePstress}. The lattice constant $c$ decreases faster than $a$ and $b$ due to the presence of vacant layers. As a result the Fe$_2$--Se bonds and the magnetic moments on Fe$_2$ atoms decrease rapidly, see Tables~\ref{tab:bondsPstress} and~\ref{tab:momentsPstress}. This difference is not so significant when the cell is compressed by small pressure. The $c$ constant changes by more than 7\% at an external pressure of more than 4.8~GPa. At the same time, the $b$ constant is slightly increased compared to the original structure. In this case, the magnetic moments on iron in both sublattices are co-directed and oppositely directed to the magnetic moments on the Se ions, which are equal to $-0.025\mu_B$ and $-0.003\mu_B$. This ferrimagnetic ordering is accomplished by the through-bond spin polarization. Iron atoms with a spin-up density induce a spin-down density on the adjacent Se atoms. This leads to an increase in the magnetic moment by $0.608\mu_B$. An increase in pressure up to 5.0~GPa leads to the disappearance of magnetic moments on Se atoms and the magnetic order of Fe$_3$Se$_4$ becomes ferromagnetic. Then, the total magnetic moment decreases with increasing external pressure that is associated with a further decrease in bond lengths, see Table~\ref{tab:bondsPstress}. At a pressure of 5.0~GPa, the value of the magnetic moment sharply decreased from $2.617\mu_B$ to $0.146\mu_B$ since the Fe$_1$--Se bond lengths starts to decrease faster than previously. The magnetic order disappears and the magnetic moment becomes zero at a pressure of 8.0~GPa. The Fe--Se bond lengths are 2.02--7.59\% shorter compared to those in the original structure. The distances between selenium atoms near the vacancy decrease by 8.9 and 6.0\%. Se--Fe$_1$--Se angles are equal to $97.7^\circ$, $91.2^\circ$, $93.5^\circ$, and $76.8^\circ$. Se--Fe$_2$--Se angles are $86.7^\circ$ and $93.3^\circ$. Selenium atoms located at opposite corners of this octahedron remain in the same plane. The volume of the cell is 201.61~\AA$^3$. Thus, the spin crossover in the monoclinic phase of Fe$_3$Se$_4$ occurs faster under compression by the isotropic pressure compared to by the isotropic strain. The dependence of the total magnetic moment under compression by an external pressure is shown in Figure~\ref{fig:DOSMagMomPstress}a.}
\begin{table}[H]
\caption{{Lattice constants (\AA) of Fe$_3$Se$_4$ under isotropic external pressure (GPa).} \label{tab:latticePstress} }
\newcolumntype{C}{>{\centering\arraybackslash}X}
\begin{tabularx}{\textwidth}{m{3cm}<{\centering}CCCCCCC}	
\toprule
& \multicolumn{7}{c}{\textbf{External Pressure (GPa)}}\\
\cline{2-8}
\raisebox {1.5ex}[0cm][0cm] {\textbf{Lattice Constant}} & \textbf{2.0} & \textbf{3.0} & \textbf{4.5} & \textbf{4.8} & \textbf{5.0} & \textbf{7.5} & \textbf{8.0}\\
\midrule
$a$ (\AA) & 6.016 & 5.993 & 5.962 & 5.883 & 5.892 & 5.857 & 5.850\\
$b$ (\AA) & 3.341 & 3.327 & 3.311 & 3.423 & 3.477 & 3.460 & 3.455\\
$c$ (\AA) & 11.065 & 11.014 & 10.929 & 10.382 & 10.048 & 10.002 & 9.987\\
\bottomrule
\end{tabularx}
\end{table}\vspace{-6pt}

\begin{table}[H]
\caption{{Bond lengths (\AA) and distance (\AA) between Se atoms closest to the iron vacancy of the Fe$_3$Se$_4$ under isotropic external pressure. The atomic numbering scheme is shown in Figure~\ref{fig:StructSpin}a.} \label{tab:bondsPstress}}
\newcolumntype{C}{>{\centering\arraybackslash}X}
\begin{tabularx}{\textwidth}{m{3cm}<{\centering}CCCCCCCC}
\toprule
\textbf{Pressure (GPa)} & \multicolumn{8}{c}{\textbf{Bond or Distance (\AA)}}\\
\midrule
2.0 & 2.440 & 2.391 & 2.371 & 2.364 & 2.529 & 2.543 & 2.936 & 5.232\\
3.0 & 2.429 & 2.382 & 2.363 & 2.360 & 2.521 & 2.528 & 2.912 & 5.214\\
4.5 & 2.414 & 2.369 & 2.350 & 2.353 & 2.511 & 2.505 & 2.878 & 5.184\\
4.8 & 2.380 & 2.339 & 2.310 & 2.337 & 2.499 & 2.432 & 2.795 & 5.071\\
5.0 & 2.377 & 2.345 & 2.307 & 2.333 & 2.457 & 2.403 & 2.772 & 4.993\\
7.5 & 2.366 & 2.332 & 2.296 & 2.450 & 2.388 & 2.736 & 2.968 & 4.968\\
8.0 & 2.362 & 2.329 & 2.293 & 2.326 & 2.447 & 2.385 & 2.730 & 4.959\\
\bottomrule
\end{tabularx}
\end{table}\vspace{-6pt}

\begin{table}[H]
\caption{{Magnetic moments (in~$\mu_B$) of iron atoms on two sites, Fe$_1$ and Fe$_2$, and total magnetic moments per Fe$_3$Se$_4$ formula unit of Fe$_3$Se$_4$ under external pressure. The Wigner--Seitz radius for Fe ions is 1.302 \AA.}  \label{tab:momentsPstress} }
\newcolumntype{C}{>{\centering\arraybackslash}X}
\begin{tabularx}{\textwidth}{CCCC}
\toprule
\textbf{Pressure (GPa)} & \textbf{Fe\boldmath{$_1$} Moment} & \textbf{Fe\boldmath{$_2$} Moment} & \textbf{Total Moment}\\
\midrule
2.0 & 1.942 & $-$1.829 & 2.097\\
3.0 & 1.884 & $-$1.738 & 2.068\\
4.5 & 1.786 & $-$1.596 & 2.009\\
4.8 & 1.065 & 0.544 & 2.617\\
5.0 & 0.077 & 0.005 & 0.146\\
7.5 & 0.013 & 0.001 & 0.017\\
8.0 & 0.000 & 0.000 & 0.000\\
\bottomrule
\end{tabularx}
\end{table}\vspace{-6pt}

\begin{figure}[H]

\includegraphics[width=0.7\linewidth]{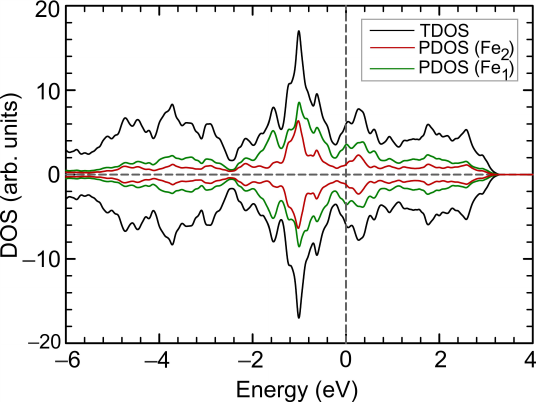}
\caption{{Total and partial DOS for Fe$_3$Se$_4$ under isotropic external pressure of 8.0 GPa. Positive and negative values correspond to spin-up and spin-down channels, respectively. The Fermi level corresponds to zero.} \label{fig:DOS8pstress}}
\end{figure}

{DOS at the Fermi level $N(E_F)$ correlates with the non-monotonic dependance of the total magnetic moment, see Figure~\ref{fig:DOSMagMomPstress}b. A decrease in the total magnetic moment is accompanied by an increase in the $N(E_F)$. A change in the magnetic order from ferrimagnetic to ferromagnetic leads to a decrease in $N(E_F)$ that was also observed in the structure under compression by the isotropic strain. The electronic structure of Fe$_3$Se$_4$ remains metallic at all values of external pressure. DOS with spin-up and spin-down become equal on each Fe cation under 8.0~GPa (Figure~\ref{fig:DOS8pstress}), resulting in a paramagnetic state.
DOS in the vicinity of the Fermi level are increased compared to the DOS of the original structure shown in Figure~\ref{fig:DOS}b. DOS redistribution in this case differs from that for the paramagnetic
Fe$_3$Se$_4$ structure at 14\% compressive strain, since the structural parameters change differently.}
	
\begin{figure}[H]

\includegraphics[width=0.7\linewidth]{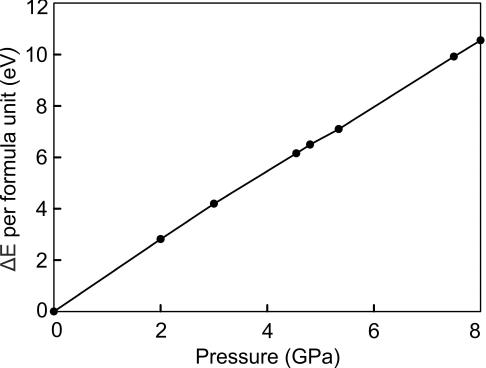}
\caption{{Energetic stability of the Fe$_3$Se$_4$ compressed by the external pressure relative to the uncompressed structure, $\Delta E$ in eV per Fe$_3$Se$_4$ formula unit.} \label{fig:EnergyPstress}}
\end{figure}

{The energy difference $\Delta E$ between compressed and uncompressed Fe$_3$Se$_4$ changes faster with the increasing external pressure than with the increasing strain, compare \mbox{Figures~\ref{fig:EnergyPstress} and~\ref{fig:Energy}}. Under the external pressure, the increase in energy is described by the linear dependence $f=1.3141x+0.1717$. Energy of Fe$_3$Se$_4$ at critical points of 5.0~GPa and 8.0~GPa is higher by $7.10$ and $10.55$~eV/f.u., respectively. Thus, the ferromagnetic structure obtained under the isotropic strain compression is more stable than the one under the isotropic external pressure by $-5.72$~eV/f.u. The energy difference of the paramagnetic states in two regimes is not so significant and equals to $-1.69$~eV/f.u.}

%%%%%%%%%%%%%%%%%%%%%%%%%%%%%%%%%%%%%%%%%%
\section{Conclusions}

\begin{figure}[H]

\includegraphics[width=0.8\linewidth]{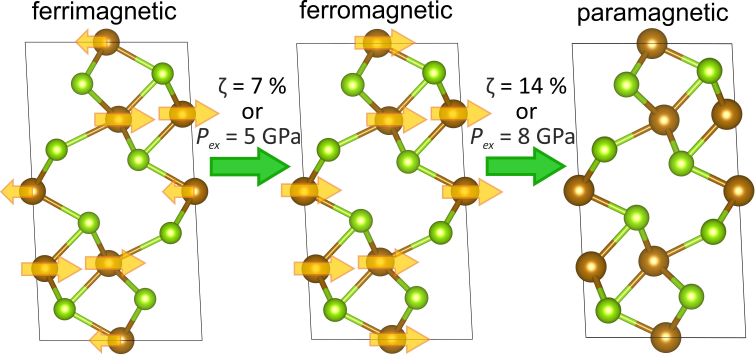}
\caption{Illustration of the Fe$_3$Se$_4$ transformation under the applied isotropic strain $\xi$ or the isotropic external pressure $P_{ex}$ between the ferrimagnetic and the paramagnetic states through the ferromagnetic state. Brown and yellow-green colors correspond to Fe and Se atoms, respectively. The arrows indicate the direction of magnetic moments on iron atoms. \label{fig:graphical}}
\end{figure}

DFT calculations within GGA for Fe$_3$Se$_4$ show that the ground state is metallic and the system is not in the half-metal state. That agrees with the conclusions of \cite{Singh2020,Persson2016}. The value of the bulk modulus was found to be 68 GPa. By studying the compression
effect on Fe$_3$Se$_4$ within DFT, we found the itinerant analogue of the spin crossover in the metallic system. In particular, we calculated the magnetic moment in each of two iron sublattices of Fe$_3$Se$_4$ in two regimes: (a) assuming an increasing isotropic compression by the compressive strain and (b) the isotropic external pressure. For crystals with such an anisotropic chemical bonding as in Fe$_3$Se$_4$, the regime (b) is more relevant to the experiments where the isotropic external pressure is applied. If the deformation of the crystal were isotropic as in the regime (a), the values of the moments would vanish for both inequivalent positions at the 14\% of strain that corresponds to the pressure of 114~GPa. On the other hand, in the regime (b) under the isotropic external pressure, the magnetic collapse is expected to occur at a much smaller value of pressure, namely, at 8~GPa.
Under the compression,
the system evolves from the uncompressed ferrimagnetic state first to the ferromagnetic state and then to the paramagnetic
state; the process is sketched in Figure~\ref{fig:graphical}.

%%%%%%%%%%%%%%%%%%%%%%%%%%%%%%%%%%%%%%%%%%
%% optional
%\supplementary{The following are available online at \linksupplementary{s1}, Figure S1: title, Table S1: title, Video S1: title.}


%%%%%%%%%%%%%%%%%%%%%%%%%%%%%%%%%%%%%%%%%%
\authorcontributions{Conceptualization, S.G.O.; calculations, L.V.B.; formal analysis, M.M.K.; writing, L.V.B., S.G.O., and M.M.K.; funding acquisition, S.G.O. All authors have read and agreed to the published version of the manuscript.}

%%%%%%%%%%%%%%%%%%%%%%%%%%%%%%%%%%%%%%%%%%
\funding{L.V.B. and S.G.O. acknowledge the support of the Russian Science Foundation (Project 18-12-00022$\Pi$).
}

\institutionalreview{{Not applicable.}} 

\informedconsent{Not applicable.} 

\dataavailability{Not applicable.} 

%%%%%%%%%%%%%%%%%%%%%%%%%%%%%%%%%%%%%%%%%%
\acknowledgments{We acknowledge the useful discussions with M.A. Vysotin. L.V.B. would like to thank the Information Technology Center, Novosibirsk State University, for providing access to their supercomputer facilities.}

%%%%%%%%%%%%%%%%%%%%%%%%%%%%%%%%%%%%%%%%%%
\conflictsofinterest{The authors declare no conflicts of interest.}

%%%%%%%%%%%%%%%%%%%%%%%%%%%%%%%%%%%%%%%%%%
\begin{adjustwidth}{-\extralength}{0cm}

%%%%%%%%%%%%%%%%%%%%%%%%%%%%%%%%%%%%%%%%%%
\reftitle{References}


\externalbibliography{no}
%\bibliography{mmkbibl9}

%=====================================
% References, variant B: internal bibliography
%=====================================

%%%%%%%%%%%%%%%%%%%%%%%%%%%%%%%%%%%%%%%%%%
\end{adjustwidth}
\end{document}